\documentclass[lettersize,journal]{IEEEtran}
\usepackage{amsmath,amsfonts,amssymb}
\usepackage{array}
\usepackage[caption=false,font=footnotesize]{subfig}
\usepackage{textcomp}
\usepackage{color}
\usepackage{stfloats}
\usepackage{url}
\usepackage{graphicx}
\usepackage{cite}
\makeatletter
\let\NAT@parse\undefined
\makeatother
\usepackage{hyperref}

\usepackage{multirow}
\usepackage{bm}

\begin{document}

\title{Channel Measurement and Characteristics Analysis for Indoor Flexible Pinching Antenna Systems in the mmWave Band}

\author{Haoyang Li, Weidong Liu, \IEEEmembership{Member,~IEEE}, Songjiang Yang, \IEEEmembership{Member,~IEEE}, Zhensheng Chen, \IEEEmembership{Member,~IEEE}, Jing Zhu, \IEEEmembership{Member,~IEEE}, Gaojie Chen, \IEEEmembership{Senior Member,~IEEE}, and Wei Huang

\thanks{This work was supported in part by the Fundamental and Interdisciplinary Disciplines Breakthrough Plan of the Ministry of Education of China under Grant JYB2025XDXM406, and by the Fundamental Research Funds for the Central Universities, Sun Yat-sen University under Grant No.24hytd010.}
\thanks{Haoyang Li, Zhensheng Chen, Jing Zhu, Gaojie Chen and Wei Huang are with the School of Flexible Electronics (SoFE), Sun Yat-sen University, Shenzhen, Guangdong 518107, China (e-mail: lihy567@mail2.sysu.edu.cn; chenzhsh27@mail.sysu.edu.cn; zhuj229@mail.sysu.edu.cn; chengj235@mail.sysu.edu.cn; huangw323@mail.sysu.edu.cn).
}
\thanks{Songjiang Yang is with the School of Communication and Information Engineering, Shanghai University, Shanghai 200444, China (e-mail:yangsongjiang@shu.edu.cn).
}
\thanks{Weidong Liu is with the School of Telecommunications Engineering, Xidian University, Xi'an 710071, China, and also with the School of Flexible Electronics (SoFE), Sun Yat-sen University, Shenzhen, Guangdong 518107, China (e-mail: wdlui@stu.xidian.edu.cn).
}
}


\maketitle

\begin{abstract}
  Pinching antenna (PA) systems offer flexible antenna configurations for future wireless communications, yet how PA geometry, orientation, and position affect channel characteristics in practical environments remains insufficiently studied. To address this gap, this paper first presents an indoor PA system channel measurement campaign at 59.5~GHz. A dedicated time-domain channel sounder is developed with frequency conversion and a circular-waveguide-to-dielectric-waveguide transition to feed the radio frequency (RF) signal into a flexible dielectric waveguide. Using this platform, office measurements quantify channel attenuation and channel characteristics by examining four PA geometries, arc PA orientation, circle PA pinching position, and transmitter--receiver height difference. The channel measurement data are analyzed in terms of path loss, shadow fading, root-mean-square (RMS) delay spread, Rician K-factor, normalized capacity, and spatial similarity. Indoor path loss models for PA systems are fitted for each PA geometry. The results reveal geometry-dependent attenuation and shadow-fading variability, with the arc PA showing greater variability than the circle PA. Arc PA rotation shifts the delay spread and conditional K-factor distributions. Moving the circle PA produces nonmonotonic changes in delay spread and normalized capacity, while spatial similarity remains relatively high at many receiver locations as the pinching position changes. These observations provide an empirical basis for PA systems channel model and system design.
\end{abstract}


\section{Introduction}
Next-generation wireless networks are expected to support higher data rates, more reliable connections, and greater device densities \cite{ma2026a,huang2026flexible}. A key approach to meeting these demands is to exploit higher-frequency spectrum, which offers access to wider bandwidths \cite{zhao2026reconfigurable}. However, links at these frequencies are more susceptible to blockage, making line-of-sight (LoS) propagation especially important. Flexible pinching antenna (PA) systems comprise low-loss dielectric waveguides that carry radio-frequency (RF) signals and dielectric elements placed near the waveguides that serve as pinching antennas, coupling part of the guided signal into free space at selected positions \cite{fukuda2022pinching}. Repositioning the PAs along the waveguides can bring the radiating points closer to users or provide less obstructed paths, improving LoS availability \cite{ding2025flexible}.

Motivated by the potential of flexible PA placement to improve LoS availability, studies have analyzed large-scale path loss, developed transmission strategies, and explored diverse applications of PA system \cite{xu2025rate,Tegos2025minimum,li2026pinching,kumar2026performance,yang2026rate,li2025total,wang2026pinching}. For downlink transmission, the authors of \cite{xu2025rate} optimized the PA locations in a single-user system by accounting for the coupled path loss and phase effects of the waveguide and free-space links, and developed a two-stage low-complexity algorithm for transmission rate maximization. For multi-user uplink transmission, the authors of \cite{Tegos2025minimum} jointly optimized the PA locations and resource allocation to maximize the minimum user rate, achieving improved fairness and performance compared with conventional fixed-antenna systems. The same flexibility in PA configuration has also been explored for integrated sensing and communication (ISAC), high-efficiency transmission, Internet of Things (IoT), and physical-layer security. In \cite{li2026pinching}, the authors proposed a multi-waveguide PA-assisted ISAC system that jointly optimizes digital beamforming and PA placement to enhance the sensing performance under communication quality of service (QoS) and power constraints, with simulations showing better sensing performance than conventional antenna schemes. The authors of \cite{kumar2026performance} derived closed-form outage expressions for communication and sensing in a PA-assisted ISAC system and showed that it improves communication reliability and sensing accuracy over conventional counterparts. In \cite{yang2026rate}, the authors optimized PA placement and backscatter reflection in an uplink IoT PA system, achieving higher rates and lower outage than fixed-antenna systems. To support energy-efficient downlink transmission, the authors of \cite{li2025total} optimized power allocation, PA activation, and PA placement to reduce total transmit power while meeting transmission rate requirements. The authors of \cite{wang2026pinching} used cooperative PA activation to strengthen the legitimate link and weaken the eavesdropper link, thereby improving the secrecy rate of PA systems. In \cite{zhao2026robust}, the authors developed a blockage-aware design for multi-waveguide PA systems that jointly optimizes PA placement, beamforming, and artificial noise under imperfect channel state information to improve user rates while satisfying secrecy constraints.

In addition to system-level optimization and applications, recent studies have examined PA radiation characteristics and their implications for PA systems design \cite{li2026geometry,li2026unlocking,zhang2026directional,zubair2026closed}. The authors of \cite{li2026geometry} fabricated a PA system prototype and provided the first experimental verification of PA radiation directivity by measuring the radiation patterns of PAs with different geometries. They further designed an arc-shaped PA and showed through simulations that rotating it can steer the radiation direction. The authors of \cite{li2026unlocking} developed a geometry-aware channel model and a PA system prototype for video transmission, demonstrating how PA shape and orientation affect directional coverage and wireless signal quality. In \cite{zhang2026directional}, the authors proposed directional PA systems, which models directional radiation, waveguide loss, and LoS blockage to guide PA placement. In \cite{zubair2026closed}, the authors derived a closed-form PA radiation-pattern model for a two-dimensional PA systems using coupled-mode theory and a radiation integral, and validated it against full-wave simulations. These studies have clarified how PA radiation affects coverage and link performance. Characterizing the PA systems channel also requires accounting for propagation along the waveguide, deployment conditions, and environmental multipath. Channel measurements are therefore needed to quantify these effects and support realistic models for PA systems design and performance evaluation.

Channel measurements of other position-reconfigurable antenna technologies have provided useful experimental references. Movable antenna (MA) and fluid antenna (FA) exploit antenna-position degrees of freedom to adapt to spatial channel variations \cite{zhu2024movable,wong2021fluid,zhu2025fluid,zhu2024modeling}. Some measurements have been conducted for these advanced antenna systems. In \cite{dong2026movable}, the authors developed and experimentally evaluated an ultra-accurate MA prototype, validating the ability of antenna-position adaptation to exploit spatial channel diversity and improve received-signal power in practical propagation environments. The authors of \cite{wang2025channel} conducted broadband channel measurements for a two-dimensional FA system, developed a measurement-based spatially correlated channel model, and demonstrated that antenna-position optimization can mitigate multipath fading and improve spectral efficiency over fixed-position antennas. 

Beyond measurements of MA and FA systems, indoor channel studies offer useful methodological guidance for characterizing PA systems in an important deployment scenario \cite{liu2026survey}. These studies have employed both frequency-domain and time-domain sounding \cite{yang2025an}. In \cite{cai2020dynamic}, the authors employed a vector network analyzer (VNA)-based sounder with a virtual uniform circular array to characterize indoor channels and develop a dynamic channel model capturing cluster evolution and birth--death behavior. The authors of \cite{zhang2025virtual} used a VNA-based virtual antenna array to characterize path loss, delay dispersion, near-field propagation, and spatial nonstationarity. The authors of \cite{yang2026isac} employed an orthogonal frequency division multiplexing (OFDM)-based time-domain sounder to characterize static and dynamic ISAC channels in multiple propagation domains. In \cite{wang2024usrp}, the authors used a universal software radio peripheral (USRP)-based sounder to characterize indoor wireless channels at 4.9~GHz, 5.2~GHz, and 5.8~GHz, focusing on path loss, material penetration, and human blockage. 

For PA channel, the authors of \cite{yang2026channel} developed a PA channel model that jointly captures deterministic in-waveguide propagation, large-scale path loss, and small-scale fading, and applied it to the design and analysis of PA-assisted index modulation. In \cite{xu2026pinching}, the authors developed a composite statistical channel model that combines distance-dependent probabilistic LoS blockage with non-line-of-sight (NLoS) scattering and optimized the PA position under signal-to-noise ratio (SNR) and outage constraints.

Previous channel measurement studies provide useful measurement approaches, but their results do not characterize the in-waveguide propagation and PA--waveguide coupling specific to PA systems. Existing PA channel studies largely rely on theoretical models with prescribed propagation assumptions and are primarily evaluated through numerical simulations. Although these models extend PA systems analysis beyond idealized LoS propagation, measurement-based evidence linking PA configuration to channel characteristics in practical environments remains limited.

This gap is particularly relevant to PA systems because reconfiguring a PA can affect both in-waveguide propagation and the subsequent wireless link. PA geometry and orientation shape the radiation pattern and thus the relative excitation of direct and reflected paths, whereas the pinching position changes both the in-waveguide path length and the radiating point relative to the receiver and surrounding scatterers. How these configuration and deployment factors affect path loss, delay dispersion, and spatial similarity in practical environments therefore requires systematic experimental characterization. Such measurements are needed to establish an empirical basis for realistic PA channel modeling and configuration design.

Motivated by the above discussion, this paper conducts a systematic indoor channel measurement for a practical PA systems operating at 59.5~GHz. To the best of the authors' knowledge, this is the first measurement-based study that jointly characterizes the effects of PA geometry, PA orientation, Rx height, and pinching position on the PA channel. The main contributions and novelties are summarized as follows:

\begin{itemize}
  \item A USRP-based time-domain channel sounder is developed for indoor PA system measurements at 59.5~GHz. To support mmWave band measurements by USRPs, the sounder incorporates an upconverter and a downconverter driven by a shared local oscillator. Converters enable USRP-generated sounding signals to be transmitted through the PA and received signals to be digitized for channel estimation. A circular-waveguide-to-dielectric-waveguide transition is designed to efficiently feed the RF signal into the dielectric waveguide. These components form a dedicated platform for measuring the PA wireless channel.

  \item PA geometry and orientation shape the radiation pattern, while changing the pinching position alters the guided-wave path length and the radiating point. To examine how these factors and the transmitter-receiver height difference affect the PA channel characterizes, an indoor PA channel measurement is conducted at multiple receiver locations. Triangle, square, circle, and arc PAs are compared. The arc PA is rotated to examine orientation effects, and the circle PA is repositioned along the waveguide to examine pinching-position effects. Receiver height is varied while transmitter height remains fixed to assess the effect of their height difference.

  \item Indoor PA channels are characterized in terms of path loss, shadow fading, root-mean-square (RMS) delay spread, Rician K-factor, normalized capacity, and spatial similarity. Comparisons across PA geometries, arc PA orientations, and receiver heights reveal differences in attenuation variability, delay dispersion, and the relative strength of the LoS component.

  \item Experiential-based path loss models are first derived for indoor PA channels. The fitted path loss exponents quantify geometry-dependent attenuation, while shadow-fading statistics characterize variability across PA geometries, arc PA orientations, and pinching positions.
\end{itemize}

The rest of this paper is organized as follows. Section II describes the PA channel measurement system and indoor measurement campaign. Section III presents the calibration and data processing procedures used to extract channel characteristics. Section IV analyzes the channel measurement data, and Section V concludes the paper.

\section{Channel Measurement Campaign}\label{section:Campaign}
In this section, we first introduce the USRP-based channel measurement system with the PA at the transmitter. Then we describe the measurement conducted in an office at 59.5 GHz.

\begin{figure} [t!]
  \centering
  \includegraphics[width= 3.4in]{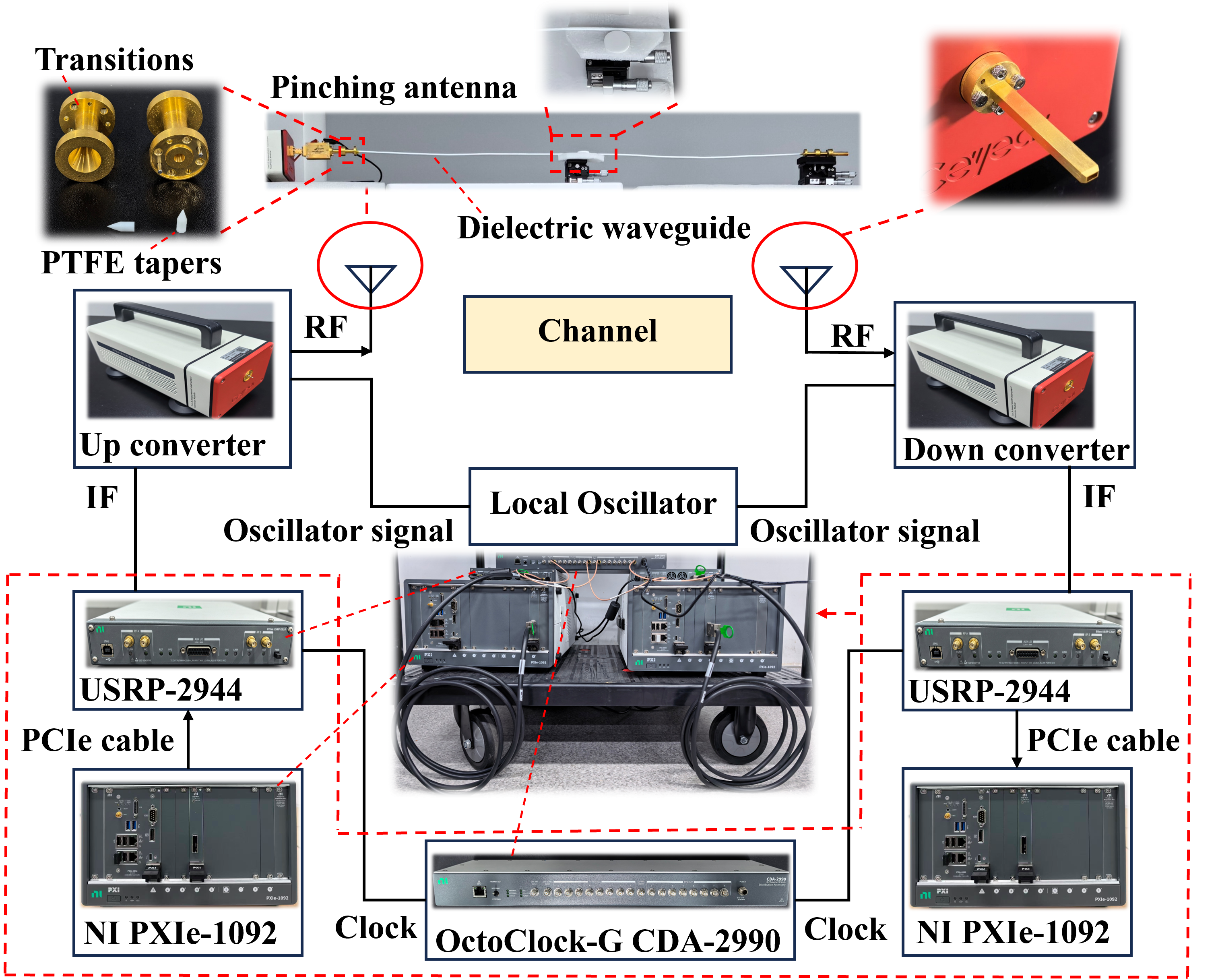}
  \caption{The diagram of the USRP-based channel measurement system for the PA system.}\label{measure_sys.Fig}
  \vspace{-1mm}
\end{figure}

\subsection{Measurement Instrumentation}
As illustrated in Fig.~\ref{measure_sys.Fig}, the transmitter (Tx) comprises an NI PXIe-1092 connected to a USRP-2944 via a PCIe cable, an upconverter, and the PA system with a PA serving as the Tx radiator. The receiver (Rx) uses a probe antenna, a downconverter, and a USRP. Integrating the PA system with the USRP requires frequency conversion to 59.5~GHz and feeding the RF signal into the dielectric waveguide. The proposed PA system comprises a dielectric waveguide and a PA that can be positioned along it to radiate the guided RF signal into free space. The PA location is referred to as the pinching position.

For frequency conversion, the upconverter and downconverter are driven by a shared 9~GHz local oscillator (LO) signal. The upconverter multiplies the LO frequency by a factor of 6 and mixes the resulting signal with the USRP-generated intermediate-frequency (IF) signal to produce the 59.5~GHz RF signal. At the Rx, the probe antenna captures the radiated signal, which is then downconverted to IF. The USRPs at both ends are synchronized using reference signals distributed by an NI OctoClock-G CDA-2990.

Based on the prototype design in \cite{fukuda2022pinching}, the PA system uses a dielectric waveguide made of polytetrafluoroethylene (PTFE), with a diameter of 3~mm and a relative permittivity of 2.1. To feed the RF signal into this waveguide, transitions from metallic circular waveguide to dielectric waveguide and PTFE tapers are designed and fabricated based on the structure in \cite{vu2021experimental}. The transition and taper dimensions are optimized for operation at 59.5~GHz. The flexible PTFE waveguide can be routed along suitable building features, such as wall or ceiling edges, while the PA can be repositioned along the waveguide to select the radiating point.

\begin{figure} [t!]
  \centering
  \includegraphics[width= 3.3in]{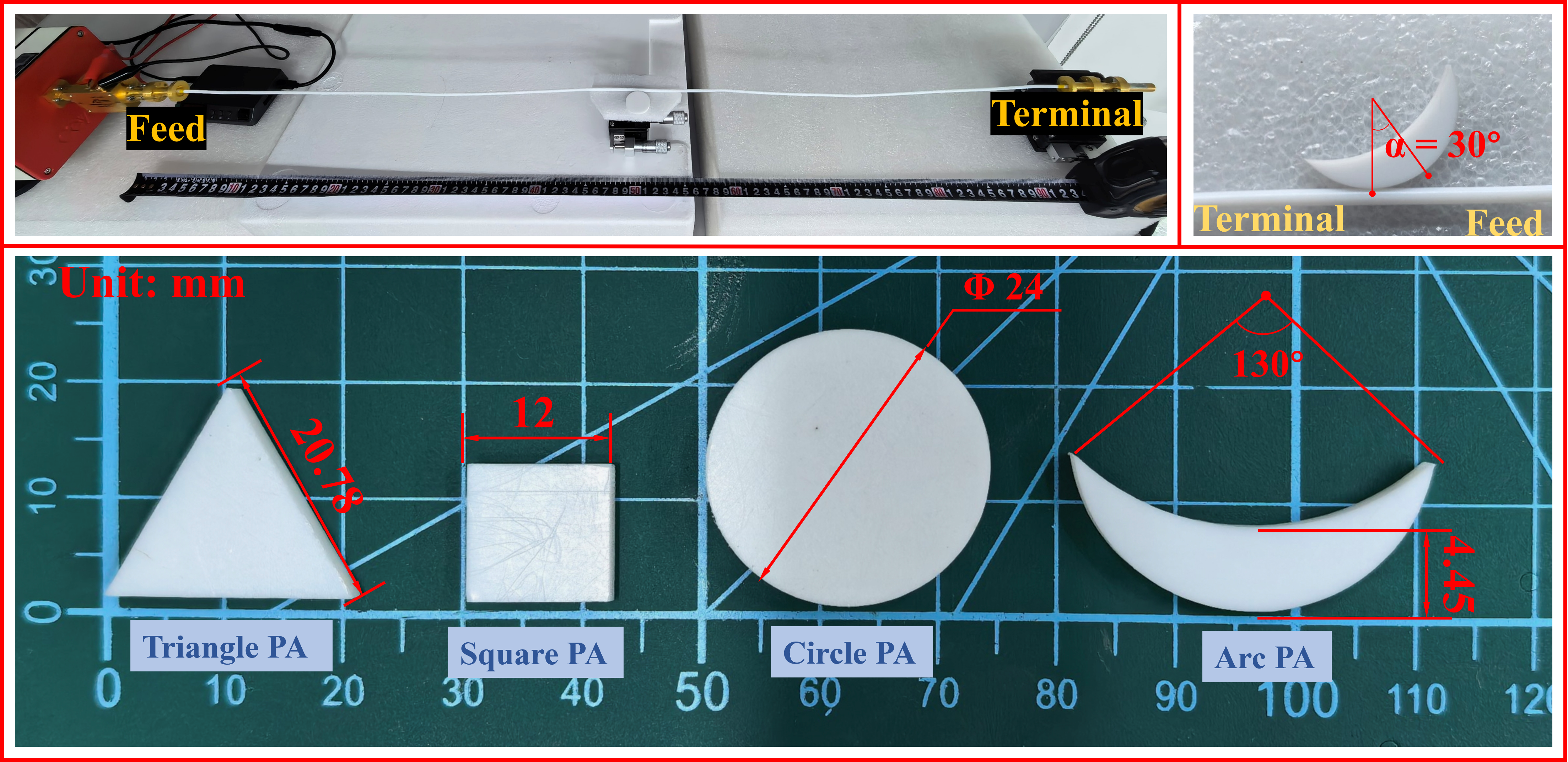}
  \caption{Fabricated PAs with different geometries.}\label{PAs.Fig}
  \vspace{-1mm}
\end{figure}

\begin{figure}[!t]
  \centering

  \subfloat[\label{fig:geometry_patterns}]
  {
      \includegraphics[width=2.5in]{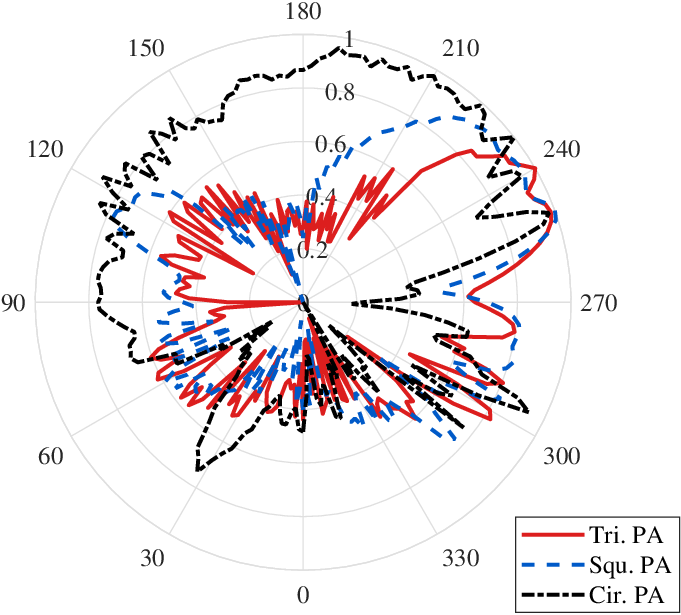}
  }
  \hfill
  \subfloat[\label{fig:arc_patterns}]
  {
      \includegraphics[width=2.5in]{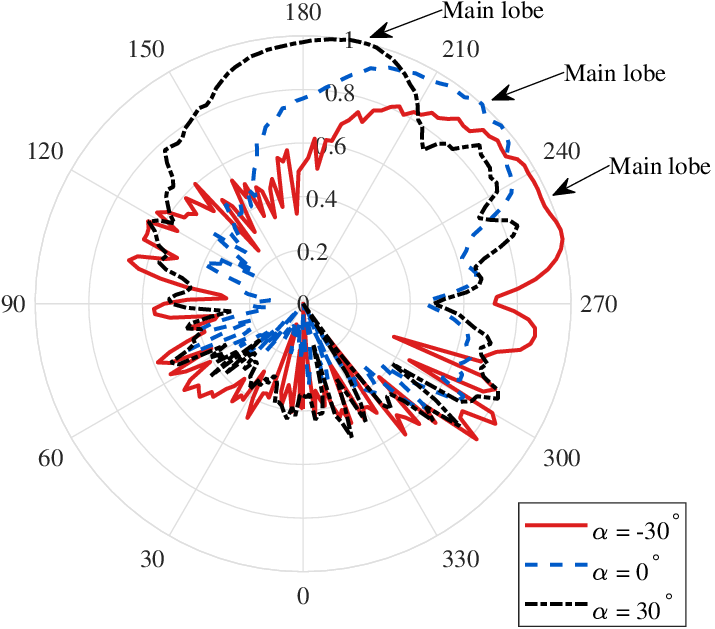}
  }

  \caption{Simulated normalized radiation patterns of the PAs used in the measurement campaign: (a) different PA geometries and (b) different rotation angles of the arc PA.}
  \label{PA_patterns.Fig}
\end{figure}

PA geometry affects the radiation pattern and hence wireless transmission. Compared with circle PAs, triangle and square PAs exhibit higher directivity. Rotating an arc PA can change its main lobe direction \cite{li2026geometry}. To examine these effects, the measurements use PAs of four geometries: triangle, square, circle, and arc. The PA geometries and dimensions are shown in Fig.~\ref{PAs.Fig}, and each PA is 3~mm thick. The normalized radiation patterns obtained from High Frequency Structure Simulator (HFSS) simulations, which broadly agree with the measurements, are shown in Fig.~\ref{PA_patterns.Fig} to illustrate the effects of PA geometry and orientation on radiation characteristics \cite{li2026geometry}.


Time-domain sounding supports repeated acquisition throughout the measurement \cite{wang2024usrp}. The downconverted IF waveform is digitized at the Rx, and the resulting complex baseband samples are stored on the host computer for offline processing. The parameters of the measurement system are shown in Table \ref{tab:measurement_parameters}.

\begin{table}[t]
  \centering
  \caption{Parameters of the Measurement System}
  \label{tab:measurement_parameters}
  \renewcommand{\arraystretch}{1.08}
  \setlength{\tabcolsep}{5pt}
  \begin{tabular}{
      >{\raggedright\arraybackslash}p{0.43\columnwidth}
      >{\centering\arraybackslash}p{0.47\columnwidth}}
      \hline
      \textbf{Parameter} & \textbf{Value} \\
      \hline
      Carrier frequency          & 59.5 GHz \\
      Transmitter height         & 2.10 m \\
      Receiver height            & 1.97 m, 1.72 m \\
      Signal bandwidth           & 80 MHz \\
      Transmit PA geometry       & Tri., Squ., Cir and Arc PA \\
      Rx 3~dB beamwidth & $115^\circ$ (E-plane), $60^\circ$ (H-plane) \\
      Rx antenna gain         & 5.42 dBi \\
      \hline
  \end{tabular}
\end{table}

\begin{figure*}[!t]
  \centering
  \includegraphics[width=\textwidth]{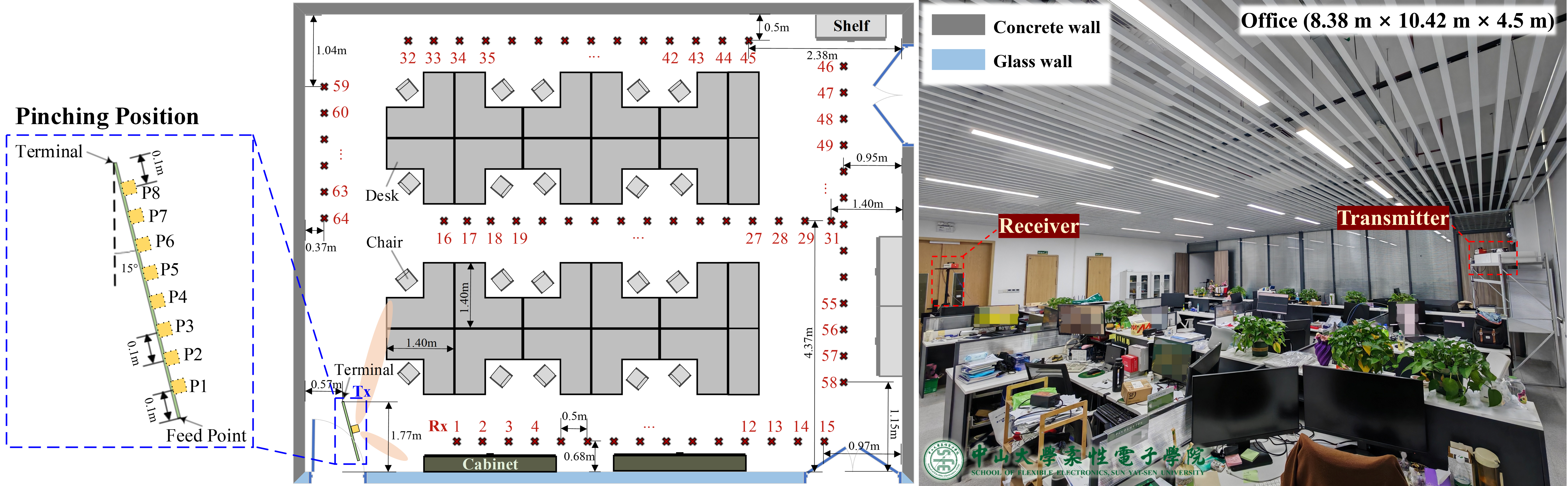}
  \caption{Office measurement environment: floor plan showing the Tx and Rx locations and an inset of the pinching positions (left), and a photograph of the office (right).}\label{environment.Fig}
\end{figure*}

\subsection{Measurement Scenario and Setup}
All measurements are carried out in an office environment, with its structure planar graph, measurement deployment, and office environment as shown in Fig. \ref{environment.Fig}. The office is bounded by concrete and glass walls and contains rows of workstations, chairs, cabinets, computer monitors, and other typical office furnishings, which provide multiple reflecting and scattering surfaces. In Fig.~\ref{environment.Fig}, the squares mark the PA pinching positions along the waveguide, while the red crosses mark the 64 horizontal Rx positions. A measurement case is defined by the PA geometry, pinching position, Rx height, and Rx position, together with the rotation angle when the arc PA is used. Adjacent Rx positions along each measurement route are separated by 0.5~m, whereas the candidate pinching positions along the 0.9~m dielectric waveguide are spaced at 0.1~m intervals. The Tx is deployed at a height of 2.10~m, above the surrounding office furniture. The Rx antenna heights used in the campaign are 1.97 m and 1.72~m.

With the office deployment and measurement positions defined above, the campaign is organized into three measurement items to examine the effects of PA geometry and Rx height, PA rotation, and pinching position:

\begin{itemize}
  \item \textbf{PA geometry and Rx height (Item 1):} The geometry of a PA can substantially affect its radiation characteristics \cite{li2026geometry}. To examine the resulting differences in the wireless channels, the PA is fixed at pinching position P5 along the dielectric waveguide, and the Rx antenna is first placed at a height of 1.97~m. The circle and arc PAs are measured at all 64 Rx positions, with the arc PA oriented at $\alpha=0^\circ$. For the more directional triangle and square PAs, usable signals are obtained mainly along the aisle between the workstation rows and the routes near the upper and left sides of the office in Fig.~\ref{environment.Fig}. Because changing the Rx height changes its elevation angle relative to the PA and may affect the received signal, the Rx antenna is then lowered to 1.72~m for further measurements.

  \item \textbf{PA rotation (Item 2):} The PAs exhibit directional radiation patterns \cite{zhang2026directional}. For the arc PA, rotating the antenna changes its main lobe direction \cite{li2026unlocking}, as illustrated in Fig.~\ref{PAs.Fig} and Fig.~\ref{PA_patterns.Fig}(b). To examine how changes in the main lobe direction affect the channel, measurements are conducted with the arc PA at rotation angles of $\alpha=-30^\circ$, $0^\circ$, and $30^\circ$. The PA remains at P5, the Rx antenna height is fixed at 1.97 m, and all 64 Rx positions are measured at each angle.

  \item \textbf{Pinching position (Item 3):} In PA systems, the PA can be repositioned along the dielectric waveguide, changing the location from which the signal is radiated. As shown in Fig. \ref{environment.Fig}, to examine how the pinching position affects the channel, the circle PA is placed near the feed point (P1), at the waveguide center (P5), and near the termination (P8). At each pinching position, all 64 Rx positions are measured with the Rx antenna fixed at a height of 1.97~m. The PA is also placed at P2--P4, P6, and P7 for finer sampling along the waveguide, with measurements at Rx positions along the aisle between the workstation rows, i.e., Rx 16--31.
\end{itemize}

\section{Measurement Data Processing}
The channel estimates contain both the propagation and system responses. We first remove the system response and compensate for antenna gains, then use the calibrated channel transfer functions (CTF) and channel impulse responses (CIR) to calculate the channel characteristics.

\subsection{System Calibration and Channel Impulse Response}
The Tx, Rx, and surrounding environment remain stationary during each acquisition. Let $\mathbf{x}(t)$ and $\mathbf{y}(t)$ denote the transmitted and received sounding signals, respectively. The transmitted waveform is an orthogonal frequency division multiplexing (OFDM) signal carrying a Zadoff--Chu (ZC) sequence. The received complex baseband signal is modeled as
\begin{equation}
  \mathbf{y}(t)=\mathbf{x}(t)*\mathbf{g}(t)*\mathbf{h}(t),
  \label{received_signal.eq}
\end{equation}
where (*) denotes the convolution in the time domain, $\mathbf{g}(t)$ is the sounder impulse response and $\mathbf{h}(t)$ is the CIR. In frequency domain, we have
\begin{equation}
  \mathbf{Y}(f)=\mathbf{X}(f)\mathbf{G}(f)\mathbf{H}(f),
  \label{frequency_domain_signal.eq}
\end{equation}
where $\mathbf{Y}(f)$, $\mathbf{X}(f)$, $\mathbf{G}(f)$ and $\mathbf{H}(f)$ represent the frequency domain transfer functions obtained by fast Fourier transform of $\mathbf{y}(t)$, $\mathbf{x}(t)$, $\mathbf{g}(t)$ and $h(t)$, respectively.

The raw CTF $\mathbf{H}_{\mathrm{raw},p}[k,m]$ is estimated from repeated sounding periods after timing alignment, phase compensation, and coherent averaging, with $k$, $p$, and $m$ denoting the active frequency, measurement case index, and snapshot indices, respectively. The raw CTF contains both the channel response $\mathbf{H}(f)$ at discrete frequencies and the measurement system response. To remove the system response and attenuation, back-to-back calibration is performed by connecting the Tx and Rx RF front ends through an attenuator under the same measurement settings. The calibration CTF is
\begin{equation}
  \mathbf{H}_{\mathrm{cal}}[k]
  =\mathbf{H}_{\mathrm{sys}}[k]G_{\mathrm{att}},
  \label{system_calibration.eq}
\end{equation}
where $\mathbf{H}_{\mathrm{sys}}[k]$ is the value of $\mathbf{G}(f)$ at the $k$th active frequency, and $G_{\mathrm{att}}$ is the known linear amplitude response of attenuator.

Using this calibration response and compensating the antenna gains, the propagation CTF is recovered as
\begin{equation}
  \widehat{\mathbf{H}}_{p}[k,m]
  =\frac{\mathbf{H}_{\mathrm{raw},p}[k,m]}{\mathbf{H}_{\mathrm{cal}}[k]}
  \frac{G_{\mathrm{att}}}
  {10^{(G_{\mathrm{Tx},p}+G_{\mathrm{Rx}})/20}}.
  \label{calibrated_ctf.eq}
\end{equation}
The Tx antenna gain $G_{\mathrm{Tx},p}$ is obtained from simulated radiation patterns validated against measurements \cite{li2026geometry}, whereas the Rx antenna gain $G_{\mathrm{Rx}}$ is taken from the measurement datasheet. Finally, the calibrated CIR is obtained by inverse fast Fourier transform
\begin{equation}
  \widehat{\mathbf{h}}_{p}[\ell,m]
  =\operatorname{IFFT}
  \left\{\widehat{\mathbf{H}}_{p}[k,m]\right\},
  \label{calibrated_cir.eq}
\end{equation}
where $\ell$ indexes the discrete delay samples of the CIR.

\subsection{Channel Characteristics}
1) \emph{Path Loss and Shadow Fading}: The path loss for each valid measurement case is calculated from the calibrated CTF. The channel power is averaged over the effective subcarriers to obtain the path loss estimate. The distance between the PA at the selected pinching position and the Rx is denoted by $d$.

The floating-intercept (FI) and close-in (CI) models are employed to characterize the distance-dependent path loss \cite{wang2024usrp,zhang2025virtual}, which are expressed as
\begin{equation}
  PL_{\mathrm{FI}}(d)
  =\beta+10\alpha\log_{10}\left(\frac{d}{d_0}\right)
  +X_{\sigma,\mathrm{FI}},
  \label{fi_path_loss.eq}
\end{equation}
and
\begin{equation}
  PL_{\mathrm{CI}}(d)
  =\mathrm{FSPL}(f_c,d_0)
  +10n\log_{10}\left(\frac{d}{d_0}\right)
  +X_{\sigma,\mathrm{CI}},
  \label{ci_path_loss.eq}
\end{equation}
respectively, where $PL_{\mathrm{FI}}(d)$ and $PL_{\mathrm{CI}}(d)$ are expressed in dB, and $d_0=1$~m is the reference distance. In the FI model, $\beta$ is the fitted intercept in dB and $\alpha$ is the path loss exponent. In the CI model, $n$ is the path loss exponent, $f_c$ is the carrier frequency, while the intercept is fixed at the free-space path loss $\mathrm{FSPL}(f_c,d_0)$. The terms $X_{\sigma,\mathrm{FI}}$ and $X_{\sigma,\mathrm{CI}}$ are the corresponding shadow-fading residuals in dB. The free-space reference loss is
\begin{equation}
  \mathrm{FSPL}(f_c,d_0)
  =20\log_{10}\left(\frac{4\pi f_c d_0}{\mathrm{c}}\right),
\end{equation}
where $\mathrm{c}$ is the speed of light. The FI parameters $\alpha$ and $\beta$ and the CI exponent $n$ are estimated by ordinary least squares.

For each dataset used in a model fit, let $i=1,\ldots,N$ index its valid measurement cases. For case $i$, $d_i$ is the distance between the PA and the Rx. The shadow-fading residual for case $i$ is calculated as
\begin{equation}
  X_{\sigma,M,i}
  =PL_i-\widehat{PL}_{M}(d_i),
  \qquad M\in\{\mathrm{FI},\mathrm{CI}\},
  \label{shadow_fading_residual.eq}
\end{equation}
where $PL_i$ is the path loss estimate for case $i$ in dB, and $\widehat{PL}_{M}(d_i)$ is the path loss predicted at $d_i$ by the fitted model $M$. The shadow-fading residual is calculated as
\begin{equation}
  \begin{aligned}
    \widehat{\sigma}_{M}
    &=
    \left[
      \frac{1}{N-p_M}
      \sum_{i=1}^{N}X_{\sigma,M,i}^{2}
    \right]^{1/2},
  \end{aligned}
  \label{shadow_fading_statistics1.eq}
\end{equation}
and
\begin{equation}
  \begin{aligned}
    \overline{X}_{\sigma,M}
    &=
    \frac{1}{N}
    \sum_{i=1}^{N}X_{\sigma,M,i},
  \end{aligned}
  \label{shadow_fading_statistics2.eq}
\end{equation}
where $p_{\mathrm{FI}}=2$ and $p_{\mathrm{CI}}=1$ are the numbers of fitted parameters, so $N-p_M$ is the residual degrees of freedom.

The goodness of fit is quantified by the coefficient of determination, given by
\begin{equation}
R_M^2 =
1-
\frac{\sum_{i=1}^{N}
\left[PL_i-\widehat{PL}_M(d_i)\right]^2}
{\sum_{i=1}^{N}
\left(PL_i-\overline{PL}\right)^2},
\qquad
M\in\{\mathrm{FI},\mathrm{CI}\},
\end{equation}
where $\overline{PL}$ is the mean of the measured path loss values. A larger $R_M^2$ indicates that the distance-dependent model explains a greater proportion of the measured path loss variation.

2) \emph{RMS Delay Spread}: The power-weighted mean delay and RMS delay spread are calculated from the retained power-delay profile (PDP) samples as
\begin{equation}
  \begin{aligned}
    \tau_{\mathrm{rms},i}
    &=
    \left[
      \frac{
        \sum_{l\in\mathcal{S}_i}
        P_{i,l}(\tau_{i,l}-\overline{\tau}_i)^2
      }{
        \sum_{l\in\mathcal{S}_i}P_{i,l}
      }
    \right]^{1/2},
  \end{aligned}
  \label{rms_delay_spread1.eq}
\end{equation}
and
\begin{equation}
  \begin{aligned}
    \overline{\tau}_i
    &=
    \frac{
      \sum_{l\in\mathcal{S}_i}P_{i,l}\tau_{i,l}
    }{
      \sum_{l\in\mathcal{S}_i}P_{i,l}
    },
  \end{aligned}
  \label{rms_delay_spread2.eq}
\end{equation}
where $\mathcal{S}_i$ contains the PDP samples within the analysis delay window whose observed powers satisfy the noise-margin and peak-relative dynamic-range thresholds. $P_{i,l}$ and $\tau_{i,l}$ are the power and delay of the $l$th retained sample, respectively.

3) \emph{K-Factor}: In the mmWave band, multipath propagation also have the impact on the indoor channel model \cite{liao2026channel,liu2026measurements}, which motivates us to further investigate the Rician K-factor of the proposed PA system. The K-factor is given by
\begin{equation}
  K_i=\frac{P_{i,\mathrm{LoS}}}{P_{i,\mathrm{NLoS}}},
  \label{k_factor.eq}
\end{equation}
where $P_{i,\mathrm{LoS}}$ is the power merged around the strongest detected sample within the expected LoS delay window, and $P_{i,\mathrm{NLoS}}$ is the detected power after the merged interval.

4) \emph{Normalized Capacity}: To compare the frequency-selective channel responses independently of large-scale attenuation, the normalized capacity is evaluated at a common reference SNR. Let $\Omega_i$ denote the set of effective subcarriers for measurement case $i$, and let $N_i$ be the number of subcarriers in this set. The CTF $\mathbf{H}_i(k)$ is scaled to unit average power over $\Omega_i$, yielding $\widetilde{\mathbf{H}}_i(k)$. At a common reference SNR, the normalized spectral efficiency is
\begin{equation}
  C_i
  =
  \frac{1}{N_i}
  \sum_{k\in\Omega_i}
  \log_2\mathrm{det}\left(\mathbf{I}_{M_R}+\frac{\rho}{M_T}\widetilde{\mathbf{H}}_i(k)\widetilde{\mathbf{H}}^\mathrm{H}_i(k)\right),
  \label{channel_capacity.eq}
\end{equation}
where $M_T$ and $M_R$ denote the numbers of Tx and Rx antennas, respectively, $\mathbf{I}_{M_R}$ denotes the $M_R$-dimensional identity matrix, and $\rho$ is the reference SNR.

5) \emph{Spatial Similarity}:
Repositioning the PA along the waveguide changes the radiating point and may alter the channel response at a given Rx position. To compare the channel responses at different pinching positions, P1 is used as the reference. At Rx position $r$, let $\mathbf{H}_{1,r}$ and $\mathbf{H}_{j,r}$ denote the calibrated, coherently averaged CTF vectors for pinching positions P1 and P$j$, respectively, restricted to their common effective subcarriers. Their normalized similarity is
\begin{equation}
  \eta_{j,r}
  =
  \frac{
    \left|\mathbf{H}_{1,r}^{\mathrm H}\mathbf{H}_{j,r}\right|
  }{
    \left\|\mathbf{H}_{1,r}\right\|_2
    \left\|\mathbf{H}_{j,r}\right\|_2
  },
  \qquad 0\leq\eta_{j,r}\leq1,
  \label{spatial_correlation.eq}
\end{equation}
where $\|\cdot \|_2$ denotes the Euclidean norm. The normalization removes overall gain differences, while the magnitude makes the measure insensitive to a constant phase offset between captures. According to \eqref{spatial_correlation.eq}, $\eta_{j,r}$ measures how closely the two complex CTFs align across the same subcarrier. Values near one indicate similar relative amplitudes and phases across frequency, whereas values near zero indicate nearly orthogonal CTF vectors.

\section{Results and Analysis}
The measurements comprised three items on a 64-position Rx grid. Item 1 compared four PA geometries at P5 with Rx heights of 1.97 and 1.72~m. The arc and circle PAs were surveyed across the full grid, whereas measurements with the triangle and square PAs were planned at Rx positions 1--6, 16--23, 32--43, and 59--64. Item 2 examined the arc PA at P5 with rotation angles $\alpha=-30^\circ$, $0^\circ$, and $30^\circ$ at an Rx height of 1.97~m. Item 3 surveyed the full grid with the circle PA at P1, P5, and P8 at 1.97~m. To examine channel variation with PA placement along the waveguide in greater detail, the circle PA was also measured at P2--P4, P6, and P7 for Rx positions 16--31.

\begin{table*}[!t]
  \centering
  \caption{Measurement items and planned Rx coverage.}
  \label{tab:measurement_items}
  \footnotesize
  \renewcommand{\arraystretch}{1.15}
  \setlength{\tabcolsep}{4pt}
  \begin{tabular}{@{}c
    >{\raggedright\arraybackslash}p{0.22\textwidth}
    >{\raggedright\arraybackslash}p{0.26\textwidth}
    c
    >{\raggedright\arraybackslash}p{0.3\textwidth}@{}}
  \hline
  Item & Factor examined & PA configuration & Rx height (m) & Planned Rx positions \\
  \hline
  1 & PA geometry and Rx height
    & Arc ($\alpha=0^\circ$) and circle PAs at P5
    & 1.97, 1.72
    & 1--64 at each height \\
  1 & PA geometry and Rx height
    & Triangle and square PAs at P5
    & 1.97, 1.72
    & 1--6, 16--23, 32--43, and 59--64 at each height \\
  2 & PA rotation
    & Arc PA at P5, $\alpha=-30^\circ$, $0^\circ$, and $30^\circ$
    & 1.97
    & 1--64 at each angle \\
  3 & PA position
    & Circle PA at P1, P5, and P8
    & 1.97
    & 1--64 at each PA position \\
  3 & PA position 
    & Circle PA at P2--P4, P6, and P7
    & 1.97
    & 16--31 at each PA position \\
  \hline
  \end{tabular}
  \vspace{0.5ex}
\end{table*}

\subsection{Path Loss and Shadow Fading}

\begin{table}[!t]
  \centering
  \caption{FI and CI Path Loss Model Parameters at 59.5 GHz}
  \label{tab:path_loss_parameters}
  \setlength{\tabcolsep}{3.5pt}
  \renewcommand{\arraystretch}{1.12}
  \footnotesize
  \begin{tabular}{lcccc}
    \hline
    Parameter & Tri. & Squ. & Arc & Cir. \\
    \hline
    $N$ & 35 & 51 & 243 & 255 \\
    $d$ (m) & 1.95--7.84 & 1.95--9.10 & 1.92--10.52 & 1.71--10.65 \\
    \hline
    \multicolumn{5}{l}{FI model}\\
    $\beta$ (dB) & 62.73 & 61.55 & 60.65 & 68.28 \\
    $\alpha$ & 2.16 & 2.26 & 2.17 & 1.51 \\
    $\sigma_{\mathrm{FI}}$ (dB) & 4.84 & 3.61 & 7.06 & 2.65 \\
    $R^2$ & 0.35 & 0.52 & 0.22 & 0.49 \\
    \hline
    \multicolumn{5}{l}{CI model} \\
    $n$ & 1.44 & 1.41 & 1.29 & 1.55 \\
    $\sigma_{\mathrm{CI}}$ (dB) & 4.92 & 3.85 & 7.22 & 2.64 \\
    $R^2$ & 0.31 & 0.44 & 0.18 & 0.49 \\
    \hline
  \end{tabular}
\end{table}

\begin{figure}[t]
  \centering
  \includegraphics[width= 3.5in]{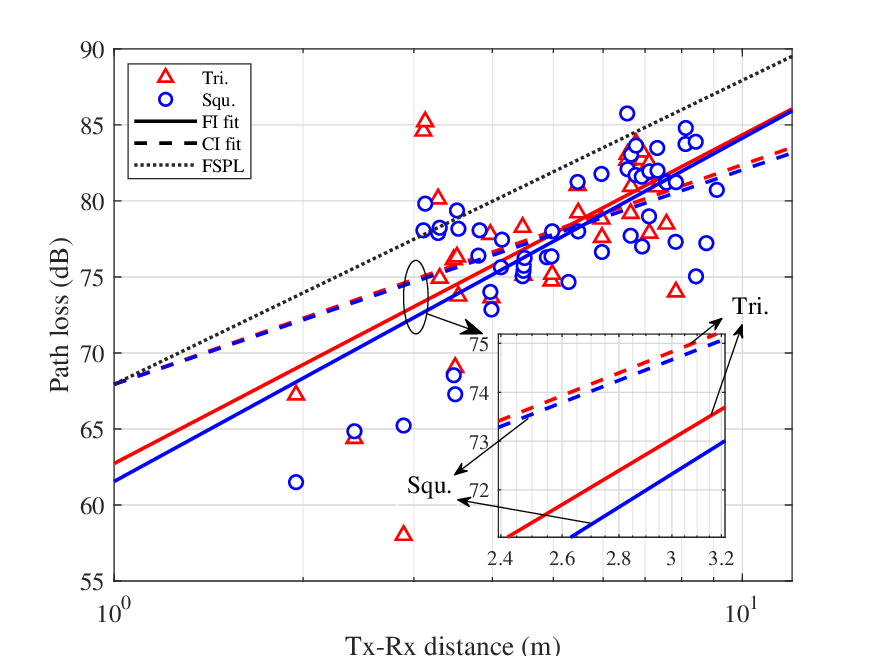}\vspace{-1em}
  \caption{Path loss of measurement results and fitted FI and CI models for the triangle and square PAs at 59.5 GHz.}
  \label{tri_squ_path loss.Fig}
\end{figure}

\begin{figure}[t]
  \centering
  \includegraphics[width= 3.5in]{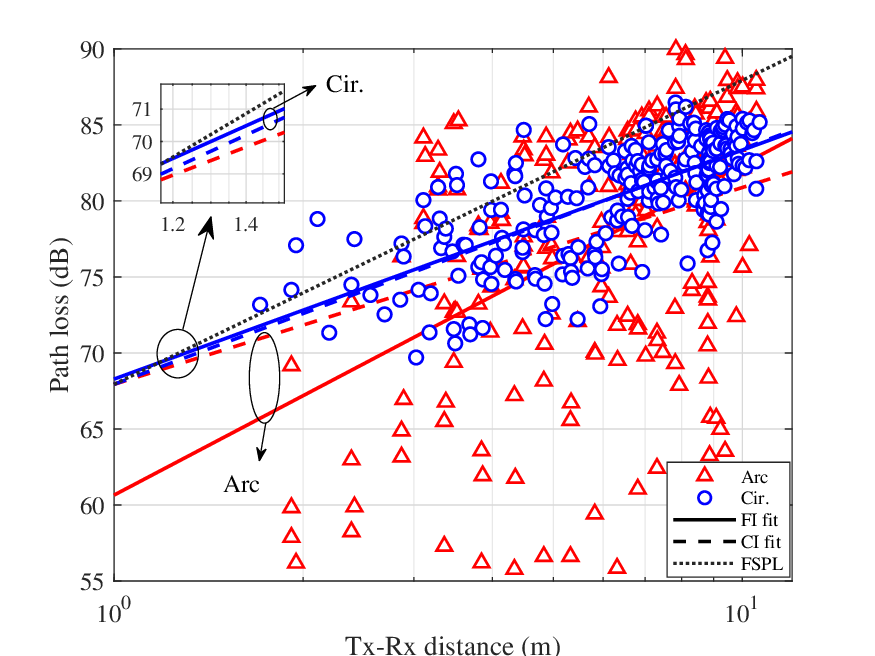}\vspace{-1em}
  \caption{Path loss of measurement results and fitted FI and CI models for the arc and circle PAs at 59.5 GHz.}
  \label{arc_cir_path loss.Fig}
\end{figure}

As some planned measurement cases did not yield usable channel data, these cases were discarded in the analysis. In Item 1, this occurred for the triangle PA at Rx positions 1--6, 19--23 and 38--43 at 1.97~m, and 4--6, 19--23 and 40--43 at 1.72~m. For the square PA, the corresponding positions were 1--6, 22, 23, 42 and 43 at 1.97~m and 4--6 at 1.72~m. Fig.~\ref{tri_squ_path loss.Fig} shows that the triangle and square PAs have similar CI exponents of 1.44 and 1.41, whereas their FI exponents are 2.16 and 2.26. Both PAs remain at P5, so the guided-wave travel distance is fixed as the Rx moves. The FI exponents, which are close to the free-space value of 2, are therefore compatible with approximately free-space-like distance decay over the retained Rx regions. Their excess over 2 may also reflect changes in obstruction and departure angle along the office routes. The narrower dominant sector of the triangle PA in Fig.~\ref{PA_patterns.Fig}(a) can weight the direct and office-reflected paths differently as the Rx moves between the workstation aisle and the wall-side routes. This provides a possible explanation for its larger CI residual standard deviation of 4.92~dB, compared with 3.85~dB for the square PA.

The arc PA at $\alpha=30^\circ$ yielded no usable channel data at Rx positions 13--15, 23, 30, 43--46, 50, 54, 57, and 58. These measurement cases were therefore not included in the analysis. Fig.~\ref{arc_cir_path loss.Fig} shows that the arc PA has a CI exponent of 1.29 and an FI exponent of 2.17, whereas the corresponding exponents for the circle PA are 1.55 and 1.51. The arc PA's orientation-dependent pattern changes which direct and reflected departure directions are illuminated across the office grid. The arc PA has a CI residual standard deviation of 7.22~dB, whereas the circle PA has a value of 2.64~dB. The orientation-dependent radiation pattern of the arc PA can change the relative excitation of direct and office-reflected paths across Rx locations, making a distance-only model less predictive. The broader angular coverage of the circle PA may provide more consistent illumination across the measured routes, in line with its smaller residual spread and lower FI exponent.

The fitted parameters are summarized in Table~\ref{tab:path_loss_parameters}. Comparing the two PA groups, the triangle and square PAs show similar CI exponents and moderate shadow fading, whereas the arc and circle PAs exhibit a larger difference in channel variability. In particular, the arc PA yields the largest shadow-fading standard deviation and the lowest $R^2$ under both models, while the circle PA yields the smallest shadow-fading standard deviation. The FI and CI $R^2$ values range from 0.22 to 0.52 and from 0.18 to 0.49, respectively, indicating that propagation distance captures only part of the measured path loss variation. For all PA types, the CI exponents range from 1.29 to 1.55 and are lower than the free-space value of 2, which can be attributed to reflected and scattered components in the indoor office.

\begin{table}[!t]
  \centering
  \caption{CI Shadow-Fading Statistics at 59.5 GHz}
  \label{tab:shadow_fading_statistics}
  \setlength{\tabcolsep}{4pt}
  \renewcommand{\arraystretch}{1.10}
  \footnotesize
  \begin{tabular*}{\columnwidth}{@{}l@{\extracolsep{\fill}}rrr@{}}
    \hline
    Condition & $N$ & $\overline{X}_\sigma$ (dB) & $\widehat{\sigma}_{\mathrm{CI}}$ (dB) \\
    \hline
    \multicolumn{4}{l}{PA geometry, $h_{\mathrm{Rx}}=1.97$~m} \\
    Arc        & 64 & $-0.53$ & 6.25 \\
    Cir.   & 63 &  0.14   & 2.74 \\
    Squ.     & 22 & $-0.10$ & 3.01 \\
    Tri. & 15 &  0.22   & 3.02 \\
    \hline
    \multicolumn{4}{l}{PA geometry, $h_{\mathrm{Rx}}=1.72$~m} \\
    Arc        & 64 & $-0.43$ & 6.38 \\
    Cir.   & 64 & $-0.02$ & 2.56 \\
    Squ.     & 29 & $-0.47$ & 4.33 \\
    Tri. & 20 & $-0.62$ & 5.72 \\
    \hline
    \multicolumn{4}{l}{Arc rotation angle} \\
    $-30^\circ$ & 64 & $-0.18$ & 3.45 \\
    $0^\circ$   & 64 & $-0.53$ & 6.25 \\
    $30^\circ$  & 51 &  0.05   & 9.94 \\
    \hline
    \multicolumn{4}{l}{Pinching position} \\
    P1 (feed)     & 64 & $-0.05$ & 2.41 \\
    P5 (center)   & 63 &  0.14   & 2.74 \\
    P8 (terminal) & 64 & $-0.01$ & 2.72 \\
    \hline
  \end{tabular*}
\end{table}

Table~\ref{tab:shadow_fading_statistics} summarizes the CI shadow-fading statistics under different measurement configurations, including the sample size, residual mean, and standard deviation. At an Rx height of 1.97~m, the arc PA has the largest residual standard deviation of 6.25~dB, while the circle PA has the smallest value of 2.74~dB. At 1.72~m, their values remain close to those at the higher Rx position, reaching 6.38~dB and 2.56~dB, respectively. The square and triangle PAs instead show larger standard deviations of 4.33~dB and 5.72~dB at 1.72~m, compared with 3.01~dB and 3.02~dB at 1.97~m. Lowering the Rx changes its elevation relative to the PA and may change the balance of direct and furniture-reflected paths.

Across the arc-rotation conditions, the CI residual standard deviations are 3.45~dB, 6.25~dB, and 9.94~dB at $-30^\circ$, $0^\circ$, and $30^\circ$, respectively. Rotation changes the arc PA's radiation direction and can therefore alter which direct and reflected paths contribute at each Rx location. For the circle PA, the standard deviations are 2.41~dB, 2.74~dB, and 2.72~dB at P1, P5, and P8, showing a narrower range across these three pinching positions. 

\subsection{Delay Spreads}

\begin{figure}[t]
  \centering
  \subfloat[]{
  \includegraphics[width= 3.5in]{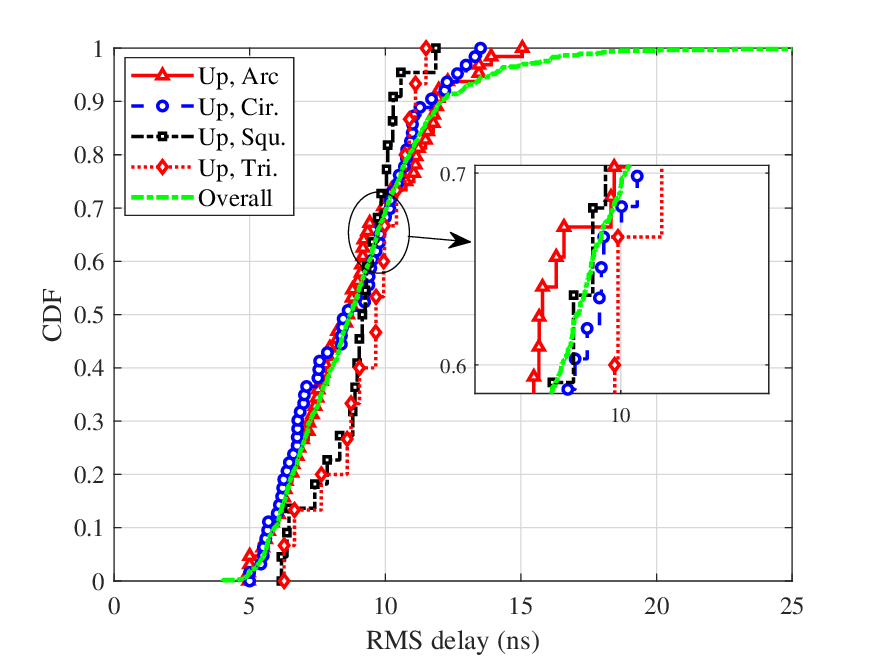}
  }
  \quad
  \subfloat[]{
  \includegraphics[width= 3.5in]{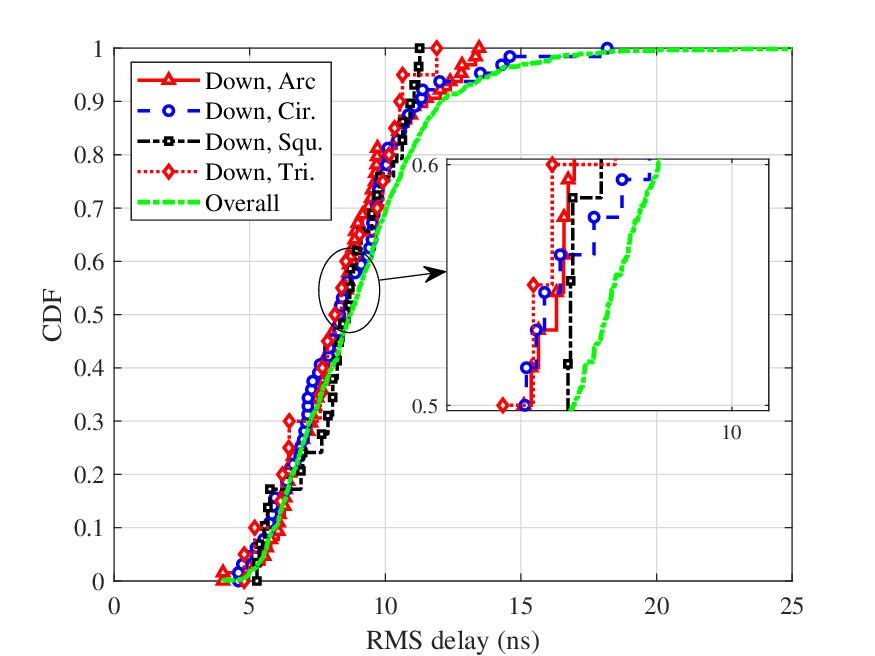}
  }
  
  \caption{CDFs of the RMS delay spread for different PA geometries at Rx heights of (a) 1.97~m and (b) 1.72~m.}\label{shape_rms.Fig}
\end{figure}

\begin{figure}[t]
  \centering
  \includegraphics[width= 3.5in]{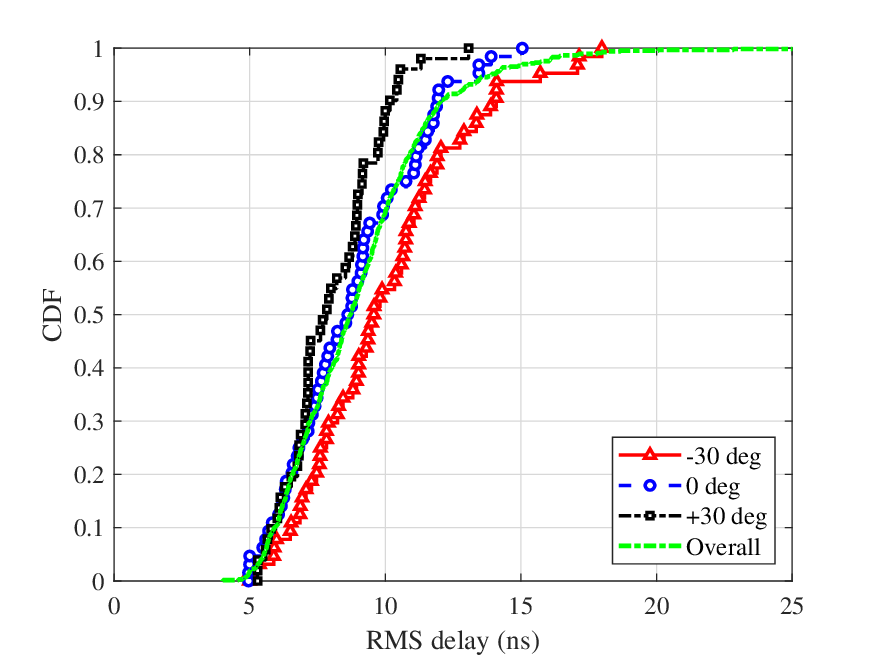}\vspace{-1em}
  \caption{CDFs of the RMS delay spread for the arc PA at different rotation angles.}
  \label{ang_rms.Fig}
\end{figure}

\begin{figure}[t]
  \centering
  \includegraphics[width= 3.5in]{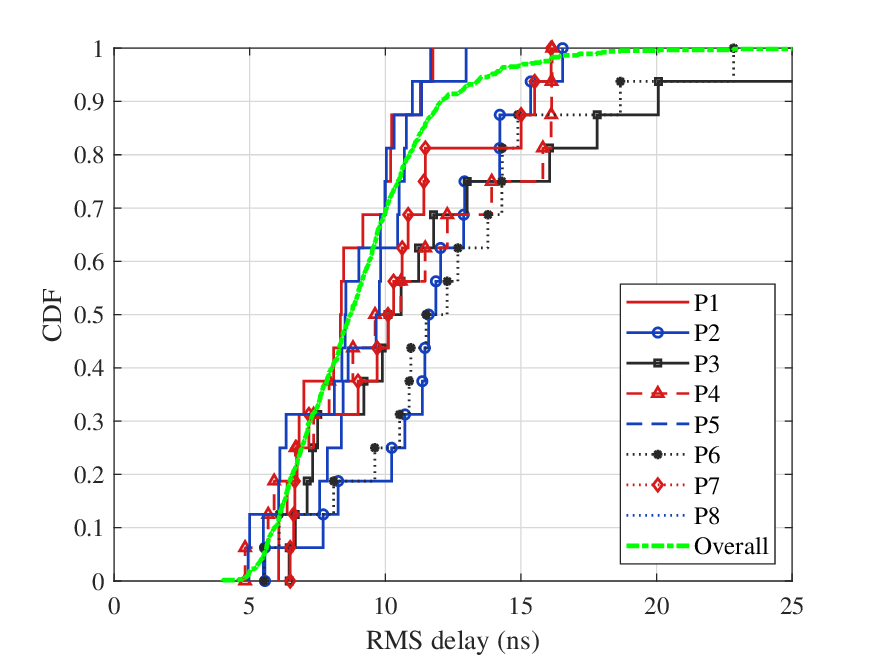}\vspace{-1em}
  \caption{CDFs of the RMS delay spread for the circle PA at pinching positions P1--P8, evaluated over the common Rx positions 16--31 ($N=16$ per pinching position).}
  \label{position_rms.Fig}
\end{figure}

Across the 664 valid measurement cases, the RMS delay spread ranges from 4.02~ns to 25.01~ns, with a mean of 8.99~ns and a standard deviation of 2.71~ns. The green dashed curve in Figs.~\ref{shape_rms.Fig}--\ref{position_rms.Fig} provides this aggregate distribution as a descriptive reference.

Fig.~\ref{shape_rms.Fig} shows substantial overlap among the central portions of the RMS delay-spread cumulative distribution functions (CDF) for the four PA geometries. At an Rx height of 1.97~m, the medians range from 8.63~ns for the circle PA to 9.66~ns for the triangle PA. At 1.72~m, they range from 8.27~ns for the triangle PA to 8.67~ns for the square PA. All four PAs are placed at P5, where changing the PA geometry can alter the relative powers of the direct and reflected paths while the office layout remains the same. The overlapping central distributions are therefore consistent with a limited geometry-dependent change in the estimated delay dispersion over their respective valid Rx positions. For the arc and circle PAs, which cover nearly the same Rx positions at both heights, the medians at 1.72~m are lower by 0.35 and 0.30~ns, respectively.

Fig.~\ref{ang_rms.Fig} shows a shift toward smaller RMS delay spreads as the arc PA is rotated from $-30^\circ$ to $30^\circ$. The medians are 9.58, 8.69, and 7.84~ns at $-30^\circ$, $0^\circ$, and $30^\circ$, respectively, while the corresponding 95th percentiles decrease from 15.47~ns to 13.28~ns and 10.52~ns. Rotation steers the arc PA's radiation pattern and changes the relative powers of the direct and delayed reflected components, which is consistent with the observed reduction in delay dispersion. 

Fig.~\ref{position_rms.Fig} compares the RMS delay spreads of the circle PA at P1--P8 over the same 16 Rx positions. The median values are lowest at P1 and P8, at 8.36~ns and 8.53~ns, respectively, and highest at P2 and P6, at 11.73~ns and 11.89~ns. The separated peaks at P2 and P6 indicate a nonmonotonic dependence on pinching position. Moving the PA changes the radiating point relative to the Rx and surrounding scatterers, which can alter the excess delays and relative powers of the received paths.

\subsection{Rician K-Factor}

\begin{figure}[!t]
  \centering
  \subfloat[]{
  \includegraphics[width= 3.5in]{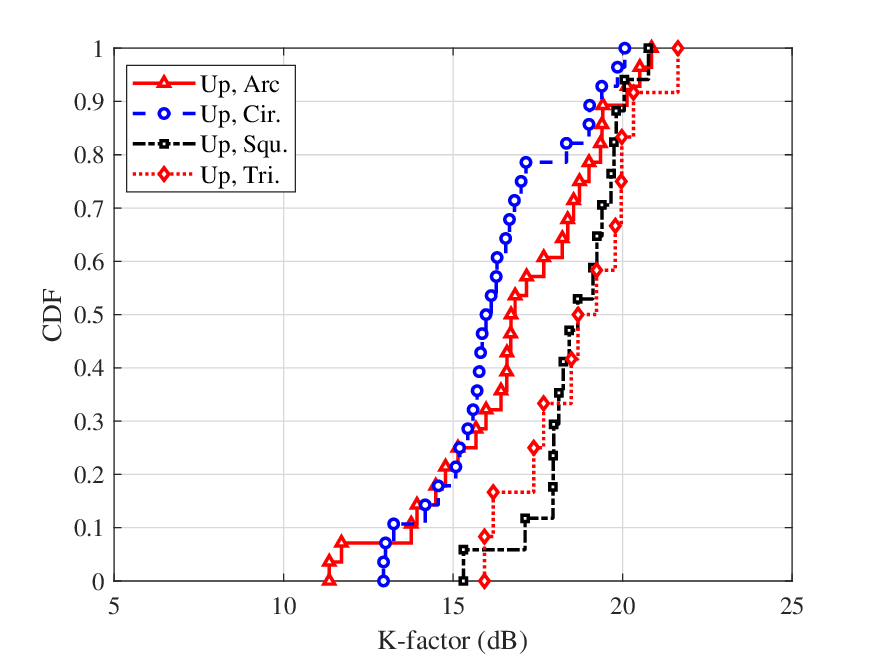}
  }
  \quad
  \subfloat[]{
  \includegraphics[width= 3.5in]{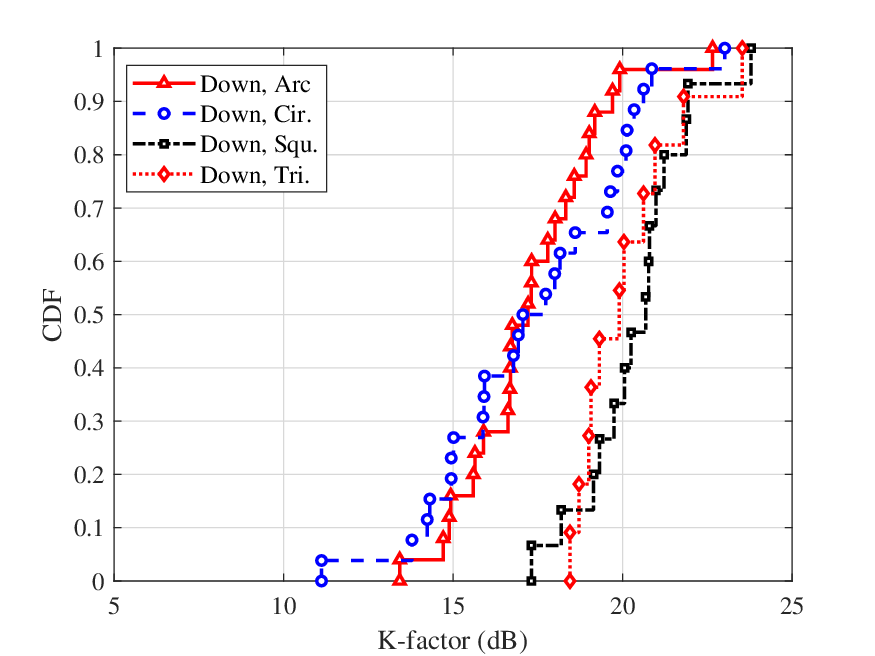}
  }
  
  \caption{CDFs of the K-factor for different PA geometries at Rx heights of (a) 1.97~m and (b) 1.72~m.}\label{shape_KF.Fig}
\end{figure}

For measurement cases with resolvable post-LoS components, Fig.~\ref{shape_KF.Fig} compares the K-factor distributions of the four PA geometries. At an Rx height of 1.97~m, the median K-factors are 16.77~dB, 16.05~dB, 18.67~dB, and 18.96~dB for the arc, circle, square, and triangle PAs, respectively. At 1.72~m, the corresponding medians are 17.21~dB, 17.39~dB, 20.68~dB, and 19.90~dB. The square and triangle PAs have higher medians at both heights, indicating a larger detected dominant-to-delayed power ratio in their measured Rx regions. Their more directional radiation patterns may favor the direct path toward the surveyed office routes while giving less weight to some paths reflected by workstations, walls, and glass. This is consistent with their higher K-factors. Across all measurement conditions, the K-factor increases as the Rx height decreases. This trend may be attributed to the largely unobstructed link at the lower Rx height, where the LoS component is more dominant relative to reflected components. At the higher Rx height, ceiling reflections may contribute more strongly.

\begin{figure}[t]
  \centering
  \includegraphics[width= 3.5in]{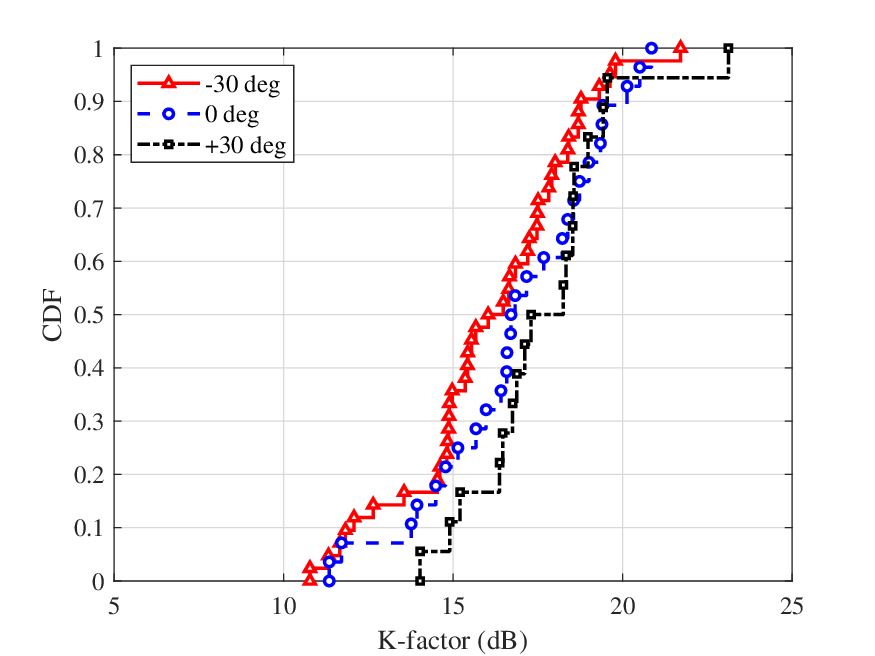}\vspace{-1em}
  \caption{CDFs of the K-factor for the arc PA at different rotation angles.}
  \label{angle_KF.Fig}
\end{figure}

Fig.~\ref{angle_KF.Fig} compares the arc PA at three orientations using eligible measurement cases. The median K-factors are 16.26~dB, 16.77~dB, and 17.78~dB for $\alpha=-30^\circ$, $0^\circ$, and $30^\circ$, respectively. The increase in these conditional medians indicates a larger dominant-to-delayed power ratio among the contributing measurement cases at $30^\circ$. As shown in Fig.~\ref{PA_patterns.Fig}(b), rotating the arc PA changes its radiation direction. In the office, the dominant path toward an Rx and delayed paths reflected by workstations, glass, and walls generally leave the PA in different directions. Rotation can therefore change their relative excitation and the resulting K-factor.

\begin{table}[!t]
  \centering
  \caption{Conditional Gaussian approximations of the Rician K-factor.}
  \label{tab:k_factor_distribution}
  \footnotesize
  \renewcommand{\arraystretch}{1.12}
  \begin{tabular*}{\columnwidth}{@{\extracolsep{\fill}}lc@{}}
  \hline
  Condition & $K_{\mathrm{dB}}$ \\
  \hline
  \multicolumn{2}{@{}l}{\emph{PA geometry, Rx height 1.97~m}} \\
  Arc      & $\mathcal{N}(16.93,\,2.48^2)$ \\
  Circle & $\mathcal{N}(16.32,\,1.93^2)$ \\
  Square   & $\mathcal{N}(18.68,\,1.29^2)$ \\
  Triangle & $\mathcal{N}(18.77,\,1.72^2)$ \\
  \hline
  \multicolumn{2}{@{}l}{\emph{PA geometry, Rx height 1.72~m}} \\
  Arc      & $\mathcal{N}(17.30,\,2.00^2)$ \\
  Circle & $\mathcal{N}(17.44,\,2.77^2)$ \\
  Square   & $\mathcal{N}(20.40,\,1.58^2)$ \\
  Triangle & $\mathcal{N}(20.12,\,1.52^2)$ \\
  \hline
  \multicolumn{2}{@{}l}{\emph{Arc PA rotation, Rx height 1.97~m}} \\
  $\alpha=-30^\circ$ & $\mathcal{N}(16.10,\,2.51^2)$ \\
  $\alpha=0^\circ$   & $\mathcal{N}(16.93,\,2.48^2)$ \\
  $\alpha=30^\circ$  & $\mathcal{N}(17.68,\,2.07^2)$ \\
  \hline
  \end{tabular*}
\end{table}

Table~\ref{tab:k_factor_distribution} summarizes descriptive Gaussian approximations to the finite K-factor samples underlying Figs.~\ref{shape_KF.Fig} and \ref{angle_KF.Fig}. In
$\mathcal{N}(\mu,\sigma^2)$, $\mu$ and $\sigma$ are the sample mean and standard deviation of $K_{\mathrm{dB}}$, respectively. Only measurement cases with resolvable multipath components contribute to these distributions, while cases without such components are omitted from the corresponding conditional CDFs. The condition $\alpha=0^\circ$ uses the same P5, 1.97~m measurements as the condition in the geometry comparison. At both Rx heights, the square and triangle PAs have higher conditional means and smaller standard deviations than the arc and circle PAs. For the arc PA, the conditional mean rises from 16.10 to 17.68~dB as $\alpha$ changes from $-30^\circ$ to $30^\circ$.

\subsection{Channel Capacity}

\begin{figure}[t]
  \centering
  \includegraphics[width= 3.5in]{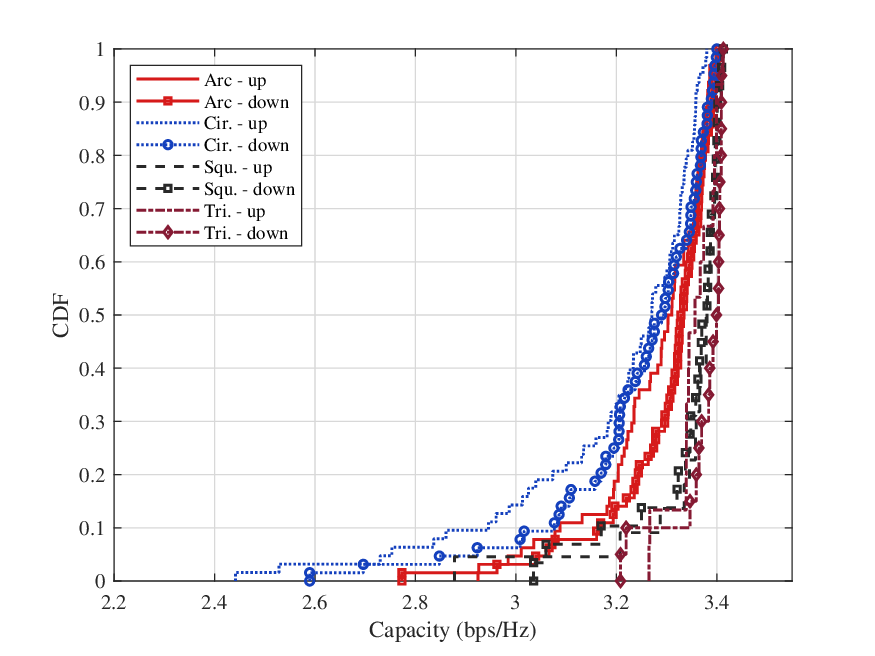}\vspace{-1em}
  \caption{CDFs of the normalized capacity for different PA geometries and Rx heights.}
  \label{shape_capa.Fig}
\end{figure}

Fig.~\ref{shape_capa.Fig} compares the normalized capacities of the four PA geometries at a reference SNR of 10~dB \cite{hou2026a}. At an Rx height of 1.97~m, the median values are 3.30~bps/Hz, 3.27~bps/Hz, 3.38~bps/Hz, and 3.36~bps/Hz for the arc, circle, square, and triangle PAs, respectively. At 1.72~m, the corresponding values are 3.33~bps/Hz, 3.29~bps/Hz, 3.38~bps/Hz, and 3.40~bps/Hz. At both heights, the square and triangle PAs lie closer to the flat-channel reference of 3.46~bps/Hz than the arc and circle PAs. The more directional square and triangle PAs may give less relative weight to some paths reflected by office furniture, walls, and glass, reducing multipath-induced spectral variation in their measured Rx regions. The circle PA's longer lower-capacity tail suggests that stronger spectral variation occurs at some of its Rx positions.

\begin{figure}[t]
  \centering
  \includegraphics[width= 3.5in]{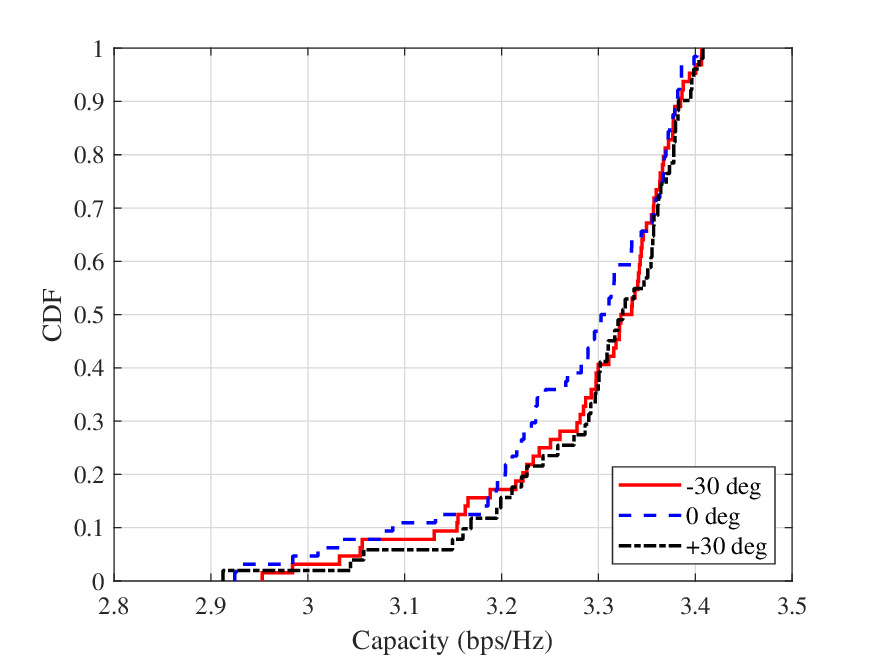}\vspace{-1em}
  \caption{CDFs of the normalized capacity for the arc PA at different rotation angles.}
  \label{angle_capa.Fig}
\end{figure}

Fig.~\ref{angle_capa.Fig} compares the normalized capacity distributions for the three arc PA orientations. Their median values are 3.32~bps/Hz, 3.30~bps/Hz, and 3.33~bps/Hz for $\alpha=-30^\circ$, $0^\circ$, and $30^\circ$, respectively. The maximum median difference of 0.03~bps/Hz indicates a small rotation effect on the spatially aggregated distribution. Rotation changes how the arc PA weights the departure directions of the dominant path and paths reflected by the office surroundings. Their altered relative powers can make the normalized channel-gain spectrum flatter at some Rx positions and more variable at others, leaving the aggregate medians close despite the changes at individual Rx positions.

\begin{figure}[t]
  \centering
  \includegraphics[width= 3.5in]{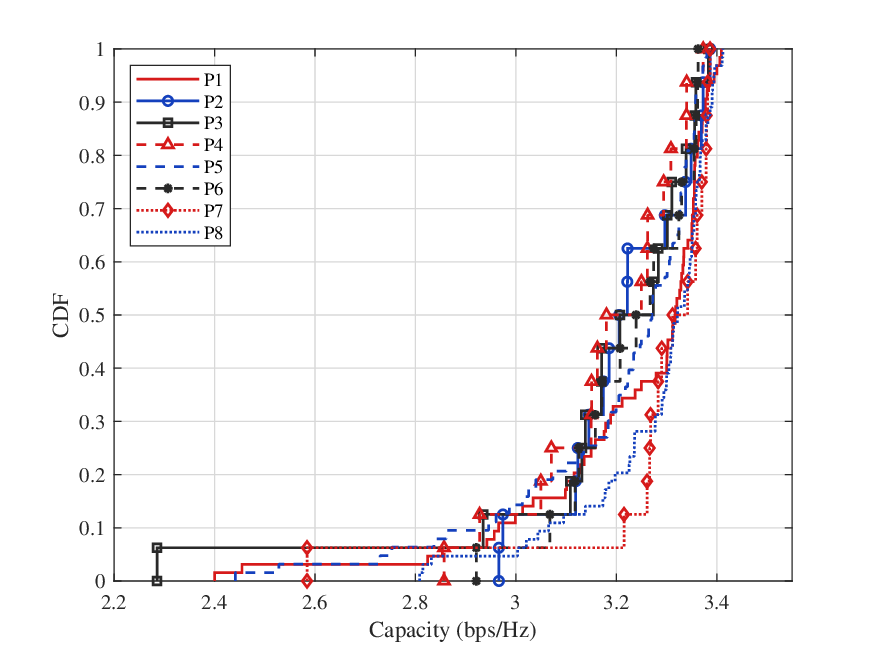}\vspace{-1em}
  \caption{CDFs of the normalized capacity for the circle PA at pinching positions P1--P8. P1, P5, and P8 include all available Rx positions, whereas P2--P4, P6, and P7 use Rx positions 16--31.}
  \label{position_capa.Fig}
\end{figure}

Fig.~\ref{position_capa.Fig} shows the normalized capacities for the circle PA at P1--P8. Since the plotted distributions use different sets of Rx positions, we additionally compare all eight pinching positions over the common Rx positions 16--31. In this matched comparison, the median capacities are 3.35~bps/Hz, 3.33~bps/Hz, and 3.31~bps/Hz at P1, P7, and P8, respectively, while the medians at P2--P6 range from 3.21~bps/Hz to 3.25~bps/Hz. The dependence on pinching position is nonmonotonic and varies across Rx locations, while the median difference between the highest and lowest capacity among the eight positions at a given Rx is 0.48~bps/Hz. Moving the PA changes the radiating point relative to the Rx and office reflectors, altering the relative path lengths, phases, and powers of direct and reflected components. The resulting changes in spectral interference are consistent with the observed position-dependent capacity and the absence of a single best position across all Rx positions.

\subsection{Spatial Similarity}

\begin{figure}[t]
  \centering
  \includegraphics[width= 3.5in]{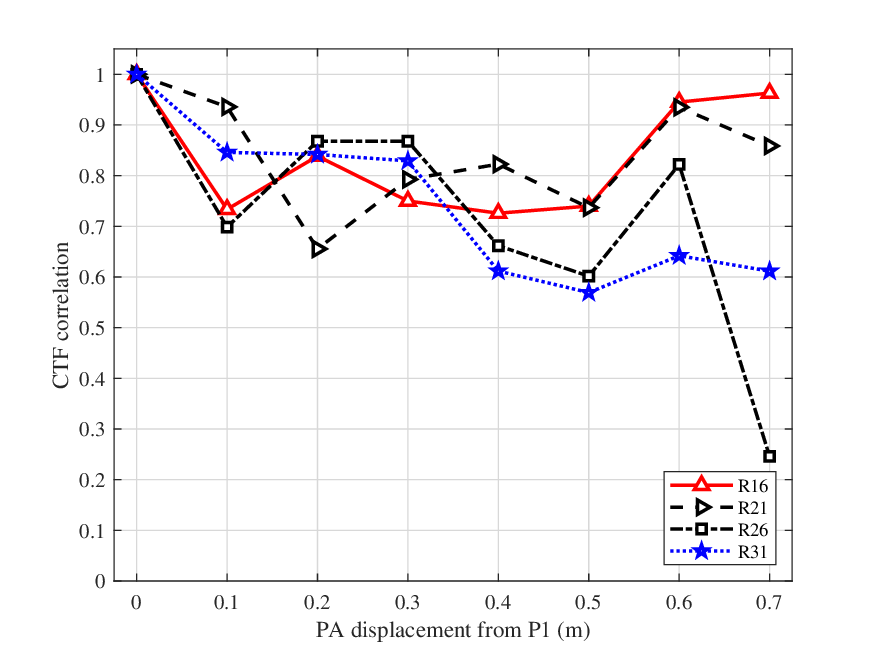}\vspace{-1em}
  \caption{Spatial similarity relative to P1 as a function of PA displacement at representative Rx positions R16, R21, R26, and R31.}
  \label{Special_correlation.Fig}
\end{figure}

Fig.~\ref{Special_correlation.Fig} shows the P1-referenced spatial similarity at four representative Rx positions. The similarity varies nonmonotonically with PA displacement and depends strongly on the Rx location. At R16, the coefficient decreases to 0.726 at P5 but rises to 0.963 at P8, whereas the coefficient at R26 falls to 0.246 at P8. Across all 16 common Rx positions, the median coefficient at P8 is 0.822, and 11 positions yield coefficients above 0.7 at the 0.7~m displacement from P1. The normalized spatial similarity thus remain similar at many Rx positions as the circle PA moves along the waveguide. Because the same PA is used within an unchanged office environment, some dominant propagation routes may continue to contribute across pinching positions. At the same time, changes in guided-wave travel distance and in the radiating point relative to the Rx and surrounding reflectors can alter path delays and relative powers, accounting for the strongly Rx-dependent results.

\section{Conclusion}\label{section:con}

This paper presented the first channel measurement for indoor PA systems at 59.5~GHz. A dedicated USRP-based time-domain sounder with a circular-waveguide-to-dielectric-waveguide transition was developed to feed RF sounding signals into a flexible dielectric waveguide and enable PA channel measurements. The measurement investigated PA geometry, arc PA orientation, pinching position, and transmitter--receiver height difference, providing experimental evidence of how PA configuration and deployment affected channel attenuation and multipath characteristics. FI and CI path loss models were derived for different PA geometries and revealed geometry-dependent attenuation and shadow fading, with the arc PA exhibiting greater shadow-fading variability than the circle PA. Arc PA rotation altered delay dispersion and the relative strength of the LoS component, while spatially aggregated normalized capacity varied little across orientations. Repositioning the circle PA produced nonmonotonic changes in delay spread and normalized capacity, and spatial similarity depended strongly on receiver location. These findings provided an empirical basis for PA channel model and system design.

\bibliographystyle{IEEEtran}
\vspace{-1em}
\bibliography{refPAchannel}

\end{document}